# A Dynamic Programming Model for Determining Bidding Strategies in Sequential Auctions: Quasi-linear Utility and Budget Constraints

**Hiromitsu Hattori**[†]　　**Makoto Yokoo**[‡]
†Nagoya Institute of Technology
Nagoya, 466-8555 JAPAN
{hatto, tora}@ics.nitech.ac.jp
http://spring.ics.nitech.ac.jp/~{hatto, tora}

**Yuko Sakurai**[‡]　　**Toramatsu Shintani**[†]
‡NTT Communication Science Laboratories
Kyoto 619-0237 JAPAN
{yokoo, yuko}@cslab.kecl.ntt.co.jp
http://www.kecl.ntt.co.jp/csl/ccrg/members/{yokoo, yuko}

## Abstract

In this paper, we develop a new method for finding the optimal bidding strategy in sequential auctions, using a dynamic programming technique. The existing method assumes that the utility of a user is represented in an additive form. From this assumption, the remaining endowment of money must be explicitly represented in each state, and the calculation of the optimal bidding strategy becomes time-consuming when the initial endowment of money $m$ becomes large.

More specifically, we develop a new problem formalization whereby the utility of a user can be represented in a quasi-linear form. By assuming a quasi-linear utility, the payment can be represented as a state-transition cost. Accordingly, we can avoid explicitly representing the remaining endowment of money. Experimental evaluations show that we can obtain more than an $m$-fold speed-up in the computation time. Furthermore, we have developed a method for obtaining a semi-optimal bidding strategy under budget constraints, and have experimentally confirmed the efficacy of this method.

## 1 Introduction

Electronic Commerce (EC) has rapidly grown with the expansion of the Internet. Among these activities, auctions have recently achieved huge popularity, and commercial auction sites (e.g., eBay; Yahoo!Auctions) have been very successful and continue to expand. Various studies on Internet auctions have already been made, from theoretical studies to practical studies (Guttman, Moukas, & Maes 1998; Harkavy, Kikuchi, & Tygar 1998; Sakurai, Yokoo, & Matsubara 1999; Sandholm 1996; Wurman, Walsh, & Wellman 1998; Yokoo, Sakurai, & Matsubara 2000).

Due to the progress of Internet auctions, a user can participate in many auctions held around the world. In some cases, such a user may have complementary/substitutional preferences on multiple items. For example, in FCC spectrum auctions (Milgrom 1998), a bidder may desire licenses for adjoining regions simultaneously (i.e., these licenses are complementary), while he/she is indifferent to which particular channel he/she receives (channels are substitutional).

One method for incorporating such complementary/substitutional preferences over multiple items is to use a combinatorial auction protocol. Research on combinatorial auctions has lately attracted considerable attention (Klemperer 1999; MacKie-Mason & Varian 1994; Milgrom 1998; Varian 1995). With a combinatorial auction protocol, multiple items with interdependent values are sold simultaneously and bidders are allowed to bid on any combination of items. Therefore, combinatorial auctions tend to increase the participants' utilities and the revenue of the seller.

However, in practice, several difficulties emerge when using combinatorial auction protocols. First, combinatorial auction protocols are very different and complicated, compared with other auction protocols used in commercial auction sites. Therefore, introducing a combinatorial auction protocol requires that both sellers and bidders drastically modify their existing systems, and learn the new protocol. In addition, determining the winners and their payments in combinatorial auctions is NP-hard, and requires considerable computational efforts (Fujishima, Leyton-Brown, & Shoham 1999; Rothkopf, Pekeč, & Harstad 1998; Sandholm 1999). Furthermore, if multiple items are to be sold by different sellers at different auction sites at different points in time, bidding for these bundles is virtually impossible.

On the other hand, in *sequential auctions* (Klemperer 1999), a set of items is sold in sequence. A bidder bids for items in a specific, known order, and he/she can choose his/her bids depending on past successes/failures. We can consider that a sequential auction mechanism is more appropriate to model existing Internet auctions. An approach for finding the optimal bidding strategy is proposed in (Boutilier, Goldszmidt, & Sabata 1999), using a dynamic programming technique for sequential auctions under several assumptions. This approach, however, assumes that the utility of the user is represented in an additive form, and accordingly, the remaining endowment of money must be explicitly represented for each state considered in the dynamic programming procedure. Therefore, the larger the initial endowment of money $m$ becomes, the more time-consuming the calculation of the optimal bidding strategy gets.

In this paper, we develop a new problem formalization that can reduce the number of states by $1/m$. In this formalization, we assume that the utility of the user can be represented in a quasi-linear form (which is an important subclass of an additive form). By representing the payment of the user as a state-transition cost, we can avoid explicitly representing the remaining endowment of money in each state. Experimental evaluations show that we can obtain more than an $m$-fold speed-up in the computation time.

The assumption of a quasi-linear utility is so general that we can deal with many cases of sequential auctions. However, there exists one practically important case where the quasi-linear representation fails to formalize, i.e., the case with budget constraints. To resolve this problem, we have developed a method for obtaining a semi-optimal bidding strategy under budget constraints, and have experimentally verified the efficiency of this method.

The rest of this paper is organized as follows. In Section 2, we introduce basic terms and definitions. In Section 3, we show the problem formalization and method for finding the optimal strategy, using the dynamic programming technique described in (Boutilier, Goldszmidt, & Sabata 1999). In Section 4, we describe a new problem formalization whereby the utility of the user can be represented in a quasi-linear form, and show that we can reduce the number of states considered in the dynamic programming procedure. Furthermore, we show that we can obtain more than an $m$-fold speed-up in the computation time by experiment. In Section 5, we describe the method for obtaining a semi-optimal bidding strategy under budget constraints, and show experimental evaluations.

## 2 Basic Models

First, we introduce several terms and definitions used in this paper. We assume that there are $n$ items denoted by $r_1, r_2, \ldots, r_n$. The individual auction $A_i$ for each item $r_i$ is executed sequentially in the increasing order of $i$, and all bidders know the order in advance. To simplify the problem, we assume that $A_i$ is a first-price sealed-bid auction (Mas-Colell, Whinston, & Green 1995). In a first-price sealed-bid auction, everyone submits a bid without knowing the others' bids. The highest bidder wins and pays the amount of his/her own bid.

In this paper, we focus on a specific agent and consider a method for finding the optimal bidding strategy for the agent. For a set of items $R_s$, which is a subset of all items $R = \{r_1, r_2, \ldots, r_n\}$, we represent the valuation of the subset $R_s$ for the agent as $v(R_s)$. We assume that the agent knows the exact value of $v(R_s)$, and that it is independent of other agents' valuations. Such items (goods) are called private-value goods (Mas-Colell, Whinston, & Green 1995).

In addition, we assume that this agent has a distribution function $F_i(h)$ to predict the highest bids of other agents in $A_i$. We also assume that the distributions of the highest bids of multiple items are mutually independent. For simplicity, we assume that the agent can win in cases of ties. Therefore, when the agent bids $z$ for an item $r_i$, the probability that the agent wins the item $r_i$ is given by $F_i(z)$, and this probability is independent of the items the agent has obtained so far.

As shown in (Boutilier, Goldszmidt, & Sabata 1999), these assumptions are rather strong. In particular, we assume that other agents do not strategically change their bids considering the bid of this agent. However, we can consider that the distribution functions only reflect the subjective probabilities of the agent, and are not necessarily correct. As shown in (Boutilier, Goldszmidt, & Sabata 1999), the agent can learn/adjust these distribution functions through experience.

## 3 Dynamic Programming in an Additive Form

In (Boutilier, Goldszmidt, & Sabata 1999), a method for finding the optimal bidding strategy, using a dynamic programming technique (Bellman 1957) is proposed based on the assumption that the utility of an agent is represented in an additive form. Dynamic programming techniques have been widely used to determine the optimal strategy in Markov decision problems (Puterman 1994). A Markov decision problem is to determine a sequence of actions that can optimize the sum of rewards/costs, given a set of states, a set of available actions for each state, the transition probabilities between states, and the rewards/costs associated with state transitions.

Assuming that after sequential auctions, an agent has obtained a set of items $R_s$ and the remaining endowment of money is $d$, a utility function in an additive form can be represented as follows.

$$v(R_s) + f(d)$$

where $f$ is some function attaching a utility to the remaining endowment of money.

To find the optimal bidding strategy, the auction process is divided into $n + 1$ stages, i.e., $n$ stages at which bidding decisions must be made, and a terminal stage at the end of all of the auctions. We use time index $0 \leq t \leq n$ to refer to stages, i.e., time $t$ refers to the point at which auction $A_{t+1}$ for item $r_{t+1}$ is about to begin. At time $t$, given the set of items $R_s$ obtained so far, and the remaining endowment of money $d$, let $<R_s, d>^t$ be the state.

Bidding strategy $\pi$ maps a state to a bid, i.e., for a state $<R_s, d>^t$, $\pi(<R_s, d>^t) = z$ means that according to the bidding strategy $\pi$, the agent should bid $z$ for $r_{t+1}$. Furthermore, we represent the expected utility for executing bidding strategy $\pi$ from the current state $<R_s, d>^t$ as $V^\pi(<R_s, d>^t)$.

The optimal bidding strategy is defined as follows. For a state $<R_s, d>^n$ in the terminal stage $n$, $V(<R_s, d>^n)$ is defined as $v(R_s) + f(d)$.

$$Q(<R_s, d>^t, z) = $$
$$F_{t+1}(z) \cdot V(<R_s \cup \{r_{t+1}\}, d-z>^{t+1})$$
$$+ (1 - F_{t+1}(z)) \cdot V(<R_s, d>^{t+1})$$
$$V(<R_s, d>^t) = \max_{z \leq d} Q(<R_s, d>^t, z)$$
$$\pi(<R_s, d>^t) = \arg\max_{z \leq d} Q(<R_s, d>^t, z)$$

The agent can determine the optimal bidding strategy $\pi$ using a method called *value iteration* (Puterman 1994). $V(<R_s, d>^t)$ represents the expected utility in the state $<R_s, d>^t$ when the agent uses the optimal bidding strategy $\pi$. In this formalization, we assume that the state transition reward/cost is 0.

In Figure 1 (a), we show an example of the optimal bidding strategy in the following simple problem setting. In this example, there are two items $r_1$ and $r_2$, and the initial endowment of money is 4. The valuation for the set $\{r_1, r_2\}$ is 4, and it is 0 for the other sets (all-or-nothing). The highest bids of the other agents for each item are 1 or 2 with equal probability 1/2, i.e., since we assume ties are wins, when the agent bids 2, its winning probability is 1, and if the agent bids 1, its winning probability is 1/2. We assume the utility for the remaining endowment of money $d$ in the final state is $d$ itself, i.e., $f(d) = d$.

In Figure 1(a), we show only the states that can occur when the agent takes the optimal bidding strategy. For each state, we show the expected utility $V$ and the optimal bid $\pi(\cdot)$. Each arrow between states represents a possible transition, and the value near the arrow represents the transition probability. In this strategy, the agent first bids 1 for $r_1$. The agent can obtain the item with the probability of 0.5. If the agent can obtain $r_1$, it bids 2 for $r_2$ in the next stage to make sure that it can obtain both items. In this case, its utility becomes $4 + 1 = 5$. Since having only $r_2$ is useless, if the agent cannot obtain $r_1$, it bids 0 for $r_2$, i.e., it does not participate in the auction. In this case, the utility becomes equal to the initial endowment of money 4. As a result, the expected utility becomes $1/2 \times 5 + 1/2 \times 4 = 4.5$.

In this problem formalization, the number of states at stage $t$ is given by $(2^t - 1) \times (m+1) + 1$, where $m$ is the initial endowment of money. This is because at stage $t$, there are $2^t$ possible combinations of obtained items. For each combination of items, the variation of the remaining endowment of money is $m+1$, excluding the case that the agent obtains nothing. Therefore, the total number of states is $O(m \times 2^n)$, which means that when the initial endowment of money is large, finding the optimal bidding strategy becomes time-consuming, as the number of states gets large.

One way to reduce the number of states is to determine bids using coarse units, e.g., considering only bids that are multiples of $10. However, unless other agents are also bidding in the same style, we cannot guarantee the optimality of the obtained strategy.

## 4 Dynamic Programming in a Quasi-linear Form

### 4.1 Basic Ideas

In this section, we introduce a new problem formalization whereby the utility of an agent can be represented in a quasi-linear form (Mas-Colell, Whinston, & Green 1995), and show that we can reduce the number of states considered in the dynamic programming procedure by $1/m$.

The utility of an agent is called quasi-linear if it can be represented in the following form.

$$v(R_s) - Z_{R_s}$$

Here, we represent the sum of payments for the subset $R_s$ as $Z_{R_s}$. We define the agent's utility as the difference between the sum of the valuation for the allocated items and the payment. The assumption that an agent's utility can be represented in a quasi-linear form has been widely used in many microeconomics studies (Mas-Colell, Whinston, & Green 1995).

Clearly, the quasi-linear form is one instance of the additive form. In a representation using an additive form, if we change the origin point to measure the utility to the utility of the initial endowment of money so that not participating in the auctions has 0 utility, the utility for obtaining the set of items $R_s$ by paying $Z_{R_s}$ is represented as $v(R_s) + f(m - Z_{R_s}) - f(m)$. Therefore, if we assume $f(x) = x$, a representation in an additive form becomes equivalent to that in a quasi-linear form.

The assumption that the utility of an agent can be represented in a quasi-linear form is reasonable if the payment for auctioned items is relatively small and has little impact on other items sold outside of the current auctions (more precisely, there exists no *income effect* (Mas-Colell, Whinston, & Green 1995)).

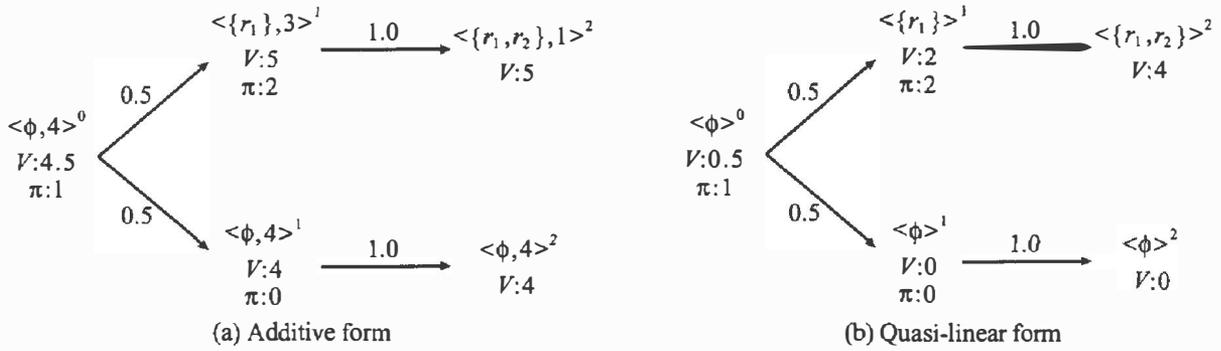

Figure 1: An Example of States and Optimal Strategy

### 4.2 Details of the Dynamic Programming Model

By assuming that the utility of an agent is represented in a quasi-linear form, and that there are no additional constraints on the amount of bids that the agent can make, the problem of finding the optimal bidding strategy, using a dynamic programming technique, can be defined as follows.

If the utility of the agent can be represented in a quasi-linear form, for two states in the terminal stage $n$, which obtain the same set of items represented as $<R_s, d>^n$ and $<R_s, d'>^n$, the difference of the utilities of these two states is identical to the difference of the remaining endowment of money $d$ and $d'$. From this fact, we can show that for two states in stage $t$, which obtain the same set of items represented as $<R_s, d>^t$ and $<R_s, d'>^t$, the optimal strategies from these states must be identical, regardless of whether the agent has obtained $R_s$ for free, or has paid a million dollars so far.

We can avoid explicitly representing the remaining endowment of money in each state by representing the payment as a state-transition cost. Let $<R_s>^t$ denote a state where an agent obtains $R_s$ at stage $t$. A bidding strategy $\pi$ maps a state to a bidding price, i.e., $\pi(<R_s>^t) = z$ means that according to bidding strategy $\pi$, the agent should bid $z$ for item $r_{t+1}$ if the agent obtains $R_s$ so far. Furthermore, let $V^\pi(<R_s>^t)$ indicate the expected utility obtained by executing strategy $\pi$ from state $<R_s>^t$. The expected utility of strategy $\pi$ of the initial state is represented as $V^\pi(<\emptyset>^0)$.

Then, we can calculate the optimal bidding strategy using value iteration in a similar way to the additive utility case. We set $V(<R_s>^n) = v(R_s)$ for state $<R_s>^n$ in the terminal stage $n$.

$$Q(<R_s>^t, z) = \\ F_{t+1}(z) \cdot (V(<R_s \cup \{r_{t+1}\}>^{t+1}) - z) \\ + (1 - F_{t+1}(z)) \cdot V(<R_s>^{t+1})$$

$$V(<R_s>^t) = \max_z Q(<R_s>^t, z)$$
$$\pi(<R_s>^t) = \arg\max_z Q(<R_s>^t, z)$$

where $V(<R_s>^t)$ denotes the expected utility of state $<R_s>^t$. In performing the value iteration, we can set the upper-bound of bidding price $z$ to $V(<R_s \cup \{r_{t+1}\}>^{t+1}) - V(<R_s>^{t+1})$.

Clearly, bidding more than this value gives a smaller expected utility than a bid of 0. On the other hand, if we assume that the utility of the agent is represented in an additive form, as discussed in (Boutilier, Goldszmidt, & Sabata 1999), there is no obvious method to set a good upper-bound of the bidding price. One obvious upper-bound of the bidding price can be the utility of having all items, if we assume $f(x) = x$. However, this upper-bound tends to be much larger than the upper-bound available in the quasi-linear representation.

Figure 1(b) shows an example of an optimal strategy in the same problem setting as Figure 1(a). As in Figure 1(a), for each state, we show the expected utility $V$ and the optimal bid $\pi$. Each arrow between states represents a possible transition, and a value near an arrow represents the transition probability.

Clearly, the optimal bidding strategies are identical for both (a) and (b). Moreover, if we recalculate the expected utility for each state in (a) by setting the valuation of the current endowment of money to the origin, the expected utilities $V$ for the corresponding states in (a) and (b) become identical. For example, in (a), the endowment of money of the initial state is 4, and $V$ is 4.5. Therefore, by participating in this auction, the agent can increase its utility (on average) by 0.5. This amount is identical to that for the initial state in (b).

In this problem formalization, the number of states at stage $t$ is given by $2^t$, and therefore, the total number of states becomes $O(2^n)$. Accordingly, compared with the case of an agent's utility represented in an additive form, the number of states in a quasi-linear form is reduced by $1/m$.

### 4.3 Evaluation

In this section, we show evaluation results to verify the efficiency of our new problem formalization.

Let $n$ be the number of items and $m$ be the initial endowment of money. We assume that $n$ is an even number, and the valuation of the set of items $r_1, r_3, \ldots, r_{n-1}$, and the set of items $r_2, r_4, \ldots, r_n$ is $100 \times n/2$. We also assume that having any additional items to these sets does not increase the utility. More specifically, if the agent has all of the items, its utility is still $100 \times n/2$ (these two sets are substitutional). If any item in each set cannot be obtained, the utility becomes 0 (i.e., the items in each set are complementary). Furthermore, we assume that the highest bids of other agents for each item are randomly distributed in the range of $[0, 100]$.

Figure 2 shows the computation time for the quasi-linear form and for the additive form, where $m$ is set to 500, 1000, and 1500, by varying the number of items $n$. In the problem formalization using the additive form, we do not consider bids larger than $100 \times n/2$. We run our experiments on a workstation (296 MHz Sun UltraSparc II) with a program written in Lisp.

We can see that in the quasi-linear form, we can reduce the number of states by $1/m$, and we can obtain more than an $m$-fold speed-up in the computation time. This is because not only the number of states, but also the number of bids considered in each state is reduced.

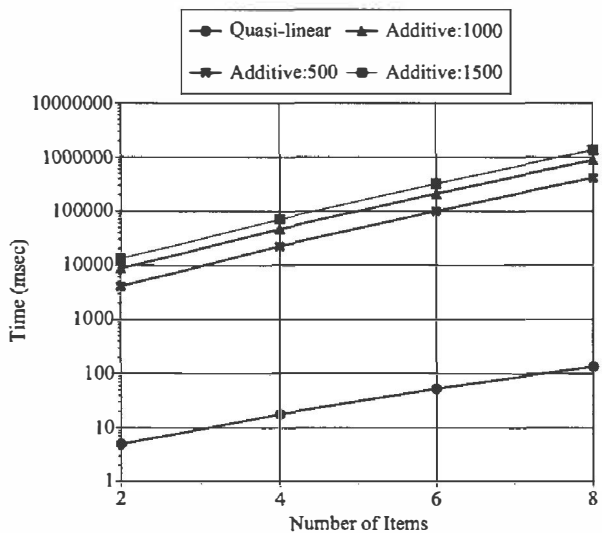

Figure 2: Comparison on Computation Time

## 5 Incorporating Budget Constraints

### 5.1 Basic Ideas

We showed that if we assume that the utility of an agent can be represented in a quasi-linear form, we can use the compact problem representation introduced in the previous section, and can reduce the number of states by $1/m$, which enables us to efficiently obtain the optimal bidding strategy by using a dynamic programming technique.

The assumption of a quasi-linear utility is so general that we can deal with many cases of sequential auctions. However, there exists one practically important case where the quasi-linear representation fails to formalize, i.e., the case with *budget constraints*. A case with budget constraints means that the sum of each bidder's payment must not exceed his/her initial endowment of money.

If we use the problem representation described in Section 3, where an agent's utility is represented in an additive form, we can obtain an optimal bidding strategy that satisfies budget constraints, because each state explicitly specifies the remaining endowment of money, and we can choose a bid so that it will not exceed the endowment of money in each state. On the other hand, the optimal bidding strategy obtained by the new problem representation might be infeasible under budget constraints, i.e., too much money might be spent in some cases.

In this section, we develop a method for obtaining a semi-optimal bidding strategy $\pi'$ that satisfies budget constraints, by modifying the strategy $\pi$ that is obtained using the method described in Section 4, i.e., by assuming that an agent's utility can be represented in a quasi-linear form, and that there are no budget constraints. More specifically, we calculate the upper-bound of a bid in each state based on the bids specified by $\pi$. Then, we find the optimal bid under this upper-bound using a dynamic programming procedure in order to satisfy the budget constraints. If we can set appropriate upper-bounds, we can find the optimal bidding strategy satisfying the budget constraints. However, because our method heuristically determines upper-bounds, we cannot guarantee the optimality of the obtained strategy.

This method applies the dynamic programming procedure twice: once for obtaining strategy $\pi$, and once for obtaining $\pi'$ by modifying $\pi$. The theoretical/experimental analysis in the previous section showed that our newly introduced problem representation gives more than an $m$-fold speed-up compared with the case of an agent's utility represented in an additive form. Therefore, we can expect our method to still attain about an $m/2$-fold speed-up in the total computation time.

## 5.2 Details of the Algorithm

As described in the previous subsection, we first obtain the optimal bidding strategy $\pi$, which does not consider budget constraints. Then, for each state, we sequentially apply a dynamic programming procedure from states in stage $n-1$.

For a state $< R_s >^t$, $Z_{pre}$ denotes the sum of the payments based on $\pi$ from the initial state $< \emptyset >^0$ to this state (excluding the payment in this state), $z_{opt}$ denotes the optimal amount of the bid in this state specified by $\pi$, and $Z_{past}$ denotes the maximal value of the sum of the payments for each of the possible paths branching from the state $< R_s \cup \{r_{t+1}\} >^{t+1}$, i.e., the state where the agent will transit if it can win item $r_{t+1}$. Note that the payments used for calculating $Z_{past}$ have already been adjusted to consider budget constraints. In addition, we denote the upper-bound of the total budget as $Z_{bud}$.

For each state $< R_s >^t$, we set the upper-bound of a bid $z_{max}$ as follows.

$$z_{max} = z_{opt} \times (Z_{bud} - Z_{past})/(Z_{pre} + z_{opt})$$

The meaning of this formula is as follows. For all states after $t + 1$, the amounts of the bids have already been adjusted. Therefore, to satisfy the budget constraints for all cases, the sum of the payments from the initial state to the state $< R_s \cup \{r_{t+1}\} >^{t+1}$ must be smaller than or equal to $Z_{bud} - Z_{past}$. The problem is how to distribute this amount among the states from $< \emptyset >^0$ to $< R_s >^t$. In this method, we simply prorate this amount based on the bids specified in $\pi$.

For all bids $z \leq z_{max}$, we calculate $Q(< R_s >^t, z)$, choose the best bid, and update $V(< R_s >^t)$.

## 5.3 Evaluation

This subsection shows experimental evaluations to confirm the efficacy of our method for finding semi-optimal strategies under budget constraints.

In Figure 3, we compare the expected utilities obtained by the method described in the previous subsection (prorated), the method in which an agent's utility is represented in an additive form (additive), and a very simple trivial method (trivial) that simply uses the optimal bidding strategy $\pi$. More specifically, in the trivial method, the agent submits bids according to $\pi$ as long as it has a sufficient endowment of money. If the remaining endowment of money becomes smaller than the amount of the optimal bid, the agent simply bids for all of the remaining endowment of money, and if the amount becomes 0, the agent stops participating in the remaining auctions.

We set the number of items to 9. For the agent, the valuation for each of the following three sets of items, i.e., $\{r_1, r_4, r_7\}$, $\{r_2, r_5, r_8\}$, and $\{r_3, r_6, r_9\}$, is 300. We assume that having any additional items except these sets does not increase the utility. More specifically, if the agent has all of the items, its utility is still 300 (these three sets are substitutional). In addition, if any item in each set cannot be obtained, the utility becomes 0. Furthermore, we assume that the highest bids of other agents for each item are randomly distributed in the range of $[0, 100]$.

Without budget constraints, the possible maximal sum of the payments for the optimal strategy is 251. In Figure 3, we show the expected utilities of the obtained strategies for three methods (additive, prorated, and trivial), by varying the budget from 10 to 260. We can see that the result of the prorated method is very close to the optimal result obtained by the additive method. On the other hand, the expected utility of the trivial method can be negative, since the agent tends to obtain only a part of the complementary items. In Figure 4, we show the computation time for these three methods. The required time for the trivial method is constant. For the prorated method, the required time for the first application of the dynamic procedure is equivalent to that for the trivial method, but the required time for the second application varies according to the budget. We can see that in the additive method, the computation time grows approximately linear to the square of the budget.

Of course, this evaluation is not extensive enough. We need to clarify the efficiency of this method in a variety of problem settings. In particular, in the example setting shown here, the winning probability changes very slowly when the agent changes its bid, and accordingly, the simple prorated method works very well. If the winning probability changes radically by a small change of the bid, it is conceivable that we will need a more sophisticated method to set the upper-bound of the bid to consider the relative impact of decreasing each bid. We are currently developing/evaluating such a method.

## 6 Conclusions

In this paper, we have proposed a method for determining the optimal strategy, using a dynamic programming technique in sequential auctions. The existing method assumes that the utility of a user is represented in an additive form, and accordingly, the remaining endowment of money must be explicitly represented in each state considered in the dynamic programming procedure. Therefore, the larger the initial endowment of money $m$ becomes, the more time-consuming the calculation of the optimal bidding strategy gets, since the number of states increases in proportion to $m$. To put it concretely, suppose that there are $n$ items and the initial endowment of money is $m$. Then the number of states considered is $O(m \times 2^n)$.

In this paper, we have developed a new problem formalization that can reduce the number of states by $1/m$. In this formalization, we assume that the utility of a user can be represented in a quasi-linear form (which is an important subclass of an additive form). By representing the payment of the user as a state-transition cost, we can avoid explicitly representing the remaining endowment of money in each state. Experimental evaluations showed that we can obtain more than an $m$-fold speed-up in the computation time. Furthermore, we have developed a method for obtaining a semi-optimal bidding strategy under budget constraints, and have experimentally confirmed the efficacy of this method.

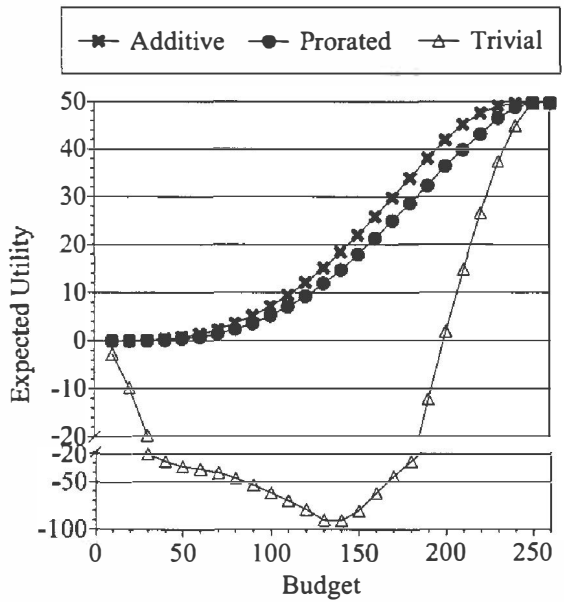

Figure 3: Comparison of the Solution Quality

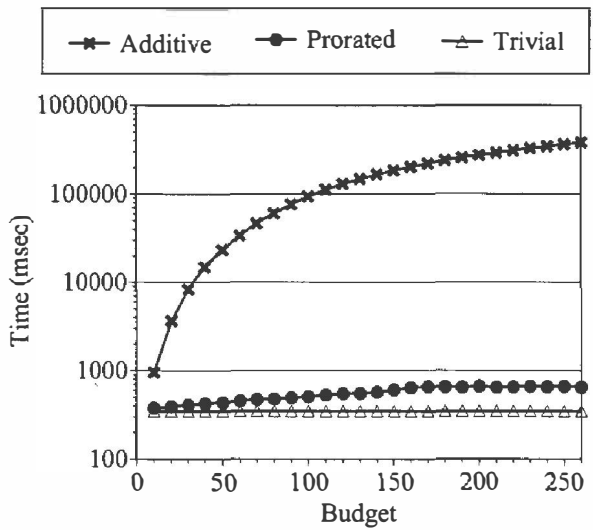

Figure 4: Comparison of the Computation Time (with Budget Constraints)

We are currently elaborating the method to find a semi-optimal bidding strategy under budget constraints, and evaluating several alternative methods in various problem settings. One future direction of this study is to develop a method for learning the optimal bidding strategy from experience, without assuming that the agent knows the distributions of the highest bids of other agents in advance. We are currently investigating a method that utilizes reinforcement learning (Barto, Bradtke, & Singh 1995) techniques.

## 7 Acknowledgment

The basic ideas for this project were developed during a visit to NTT Communication Science Laboratories by the first author. The authors wish to thank NTT for supporting the visit.